# Interdisciplinary Collaboration through Designing 3D Simulation Case Studies

Xin Bai[1]; Dana Fusco[2]

[1]Department of Teacher Education, York College of the City University of New York
[1]xbai@york.cuny.edu
[2]School of Health and Behavioral Sciences, York College of the City University of New York
[2]dfusco@york.cuny.edu

**ABSTRACT**

*Interdisciplinary collaboration is essential for the advance of research. As domain subjects become more and more specialized, researchers need to cross disciplines for insights from peers in other areas to have a broader and deeper understand of a topic at micro- and macro-levels. We developed a 3D virtual learning environment that served as a platform for faculty to plan curriculum, share educational beliefs, and conduct cross-discipline research for effective learning. Based upon the scripts designed by faculty from five disciplines, virtual doctors, nurses, or patients interact in a 3D virtual hospital. The teaching vignettes were then converted to video clips, allowing users to view, pause, replay, or comment on the videos individually or in groups. Unlike many existing platforms, we anticipated a value-added by adding a social networking capacity to this virtual environment. The focus of this paper is on the cost-efficiency and system design of the virtual learning environment.*

## 1. INTRODUCTION

Role-playing has been widely adopted as an effective pedagogical technique for students to acquire and retain knowledge and enhance skills by educators from all disciplines. It encourages learners into thinking in the subject domain as what a subject expert might think. Its advantages include: 1) it teaches empathy and understanding from different perspectives [1]; 2) it engages students in meaningful knowledge construction, which, constructivists claim, is how learning occurs [2] [3] [4]; and 3) it helps improve critical thinking and interpersonal skills. As social constructivists state, knowledge is viable not only personally, but also in social contexts [5]; students benefit through practicing in a meaningful socio-cultural context [6].

Thus, role-playing has the potential to transform theoretical concepts into an experiential format [7]. For instance, concepts learned through traditional methods such as rote, lecture, reading, writing are easily forgotten because they are often taught in isolation of direct practice; in contrast, role-playing offers educators, trainers, and learners the opportunity to observe growth and progress in both teaching and learning transactions in a more practical way [7].

In professional education, students should have plenty of opportunities to observe how domain experts (e.g., doctors, nurses, occupational therapists, social workers, teachers) make decisions and deal with complex situations based upon emergent phenomena in clinical contexts and increased opportunities to participate in the process. In real life, it is challenging to pair students with dedicated mentors due to such issues as time, costs, and safety. This is due to the issue that clinical sites are costly and difficult to obtain, Most of the time, they are reserved for the most advanced students. Thus pre-clinical "real world" scenarios would add greatly to the educational process allowing students to "see" and "participate" in clinical situations and dilemmas. Also, if faculty can control the clinical situation, professional programs would be able to better ensure and monitor the breadth of experience, e.g., working with diverse ethnic groups, in different clinical environments, with varied diagnoses and treatment approaches.

This study investigates the affordances and constraints of 3D simulation worlds with social networking capacities for encouraging multidisciplinary collaboration. We developed reusable and interdisciplinary teaching vignettes in 3D virtual learning environments. We adopted Second Life as a virtual learning tool to model clinical or hospital environments. Based upon the scripts designed by faculty from multiple disciplines, the virtual doctors, nurses, therapists, and patients interact in a computer-based virtual world. The 3D teaching vignettes were converted to video clips, allowing users to view, pause, replay, or comment on the videos individually or as groups through regular computers or mobile devices such as iPod. A technique called machinima was adopted throughout the development process. We recorded stories played out through controlling the avatars. The resulting raw video clips were edited in a process similar to how a movie is made – selecting the scenes we saw fit and adding verbal scripts on the screen as well as voiceover.

Furthermore, we embedded these teaching vignettes in a rich social networking environment through Web 2.0 tools such as Diigo (http://www.diigo.com). As more and more students are accustomed to multitasking and fast-paced communication, effective pedagogical strategies were considered to reach out to those students and engage them in meaningful learning. We documented the project development process and report findings for future directions of the study.

## 2. ROLE-PLAYING THROUGH COMPUTER-BASED SIMULATIONS

Simulation, as a teaching tool, is found to meet the broad goals within the context of any nursing program: think critically, communicate effectively, and intervene therapeutically [8]. Users can take on a new role, learn by doing and reflection, take risks, and engage in meaningful knowledge construction through proactive interaction with peers in the process. This is consistent with recent epistemologies including cognitive constructivism and social constructivism, both primarily impacting the "competent, creative, mindful, collaborative and constructive dimensions" of learning [9].

However, patient simulator mannequins usually incur a high cost. Students cannot practice anywhere anytime. Even if they can, they still lack an opportunity to observe how domain experts make decisions to deal with complex situations such as those that emerge in relation to medical crises, equipment breakdowns, "noise," or even the demanding relatives of a patient. Through computer-based 3D simulation, we could be one step closer to the educational ideal of achieving a high level of *equipment* fidelity*, environmental* fidelity*, and psychological* fidelity proposed by Beaubien and Baker [10].

Computer-based simulations have been adopted and promoted in a variety of learning contexts [11] [12] [13] [14] [15] [16] [17]. In medical and health care education, simulation methods have revolutionized training in such areas as critical care, surgery, anesthesiology, and patient safety enabling the transfer of knowledge and skills to health care providers [18] [19] [20] [21] [22].

In addition, more clinics and hospitals are expecting graduates, especially nurses, to take the leadership role of documenting the patient's story and handling electronic health records efficiently. Such demand gained momentum after the Obama administration unveiled $1.2

billion in federal grants for electronic health records systems aimed at cutting costs and improving care in the coming decade.

## 3. DESIGNING AN INTEGRATED VIRTUAL LEARNING FRAMEWORK

We developed an instructional approach that helped address the needs above in an effort to strengthen the clinical competencies of students in the allied health professions as they enter the world of work. We are guided by the broad research question: *What is the value added of integrating computer-simulation into professional education*? We were specifically interested in testing the following potential values including: cost-efficiency, student learning and motivation, and applicability to first year practice. *The focus of this paper is on the cost-efficiency and system design of the virtual learning environment.*

A virtual learning framework was developed based upon the following goals:

- Develop a virtual learning environment through interdisciplinary collaboration where students in nursing, occupational therapy, social work, and physician assistant programs can observe or role-play in a safe environment.

- Develop a research methodology to test the comparative impact of virtual and non-virtual learning environments supported by faculty in the health sciences, cognitive science, or educational technology.

We developed an immersive virtual clinical environment where students can take on the roles of doctors, nurses, patients, social workers, or even family members. Learners can observe or immerse in a 3D virtual world, understanding complex issues involved (e.g., patient symptoms, health history, insurance, religion, race) and how stakeholders (e.g., doctors, nurses, psychologist, social workers) make decisions to deal with those issues. We finished a prototype of a virtual learning environment and developed 3D teaching vignettes as videos. Those vignettes were designed by a group of seven faculty members from the following disciplines: Psychology, Occupational Therapy, Nursing, Social Work, Physician Assistant, Teacher Education, and Educational Technology. These faculty members served as domain experts and educators. They designed stories and scripts that modelled real-life scenarios to be used as teaching instruments embedded in a social-networking platform. Features such as online discussions and annotations were available in such a learning environment to prepare students to be reflective practitioners and critical thinkers able to deal with emergent situations.

Figure 1 is a screenshot of our project framework with features that allow for students to engage in role-play and collaborate with experts, instructors, and peers. Different learning skills are required when students interact in such a learning environment. For instance, practicing in a simulated real-life situation requires students to use critical-thinking and clinical-reasoning skills. Performing the scenarios in conjunction with other participants (i.e., faculty, classmates, doctors, nurses) enhances communication and delegation skills. Debriefing in class or through Web2.0 reinforces the formation of a reflective practitioner.

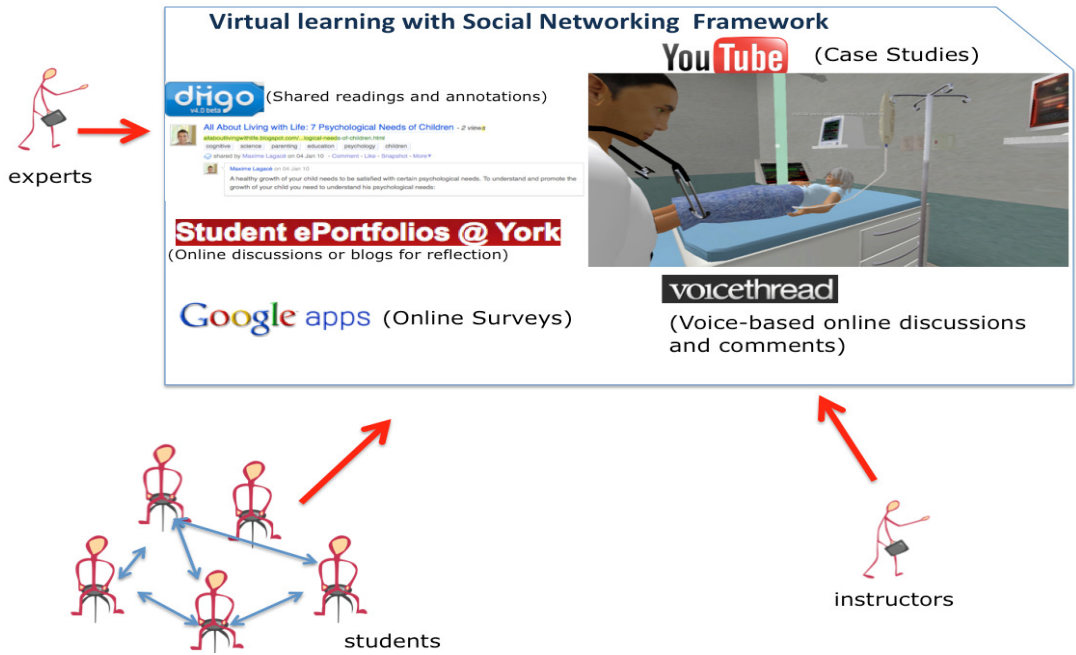

Figure 1: the virtual learning system framework

We developed our virtual learning environment in Second Life (http://secondlife.com). Second Life is a free collaborative Multi-User Virtual Environment (MUVE) whose content is generated by users. As in real life, there is a market place for virtual merchandize exchange in Second Life. Users purchase virtual land, where they can build or buy houses, trees, roads, or gardens. They can also start a business, play a game, visit a historic building or museum. More importantly, they can text or voice chat with visitors and teleport (i.e., instant transfer) to a new virtual land in a matter of click. Clubs, lecture halls, museums, or concerts become the popular gathering places of virtual users - avatars. The visual representations of avatars don't reflect the look-and-feel of real users, thus an avatar can be in the form of a human being (e.g., a doctor or a hippie), an object (e.g., a car or a tree), or an imaginary creature (e.g., a zombie or a dragon).

We developed a virtual hospital with two floors, including four clinic wards, a front desk, a cafeteria, and a waiting area. A dozen avatars in the roles of doctors, nurses, and patients were created for our students to role-play. Medical devices such as oxygen mask, blood pressure monitor, inhaler, or crane are stored in the inventory for students to retrieve. Scripts that enable such behaviours as holding hands, blowing an instrument, fainting, or washing hands were designed and embedded in the environment for users to trigger. Students can role-play in the virtual hospital anytime anywhere. For instance, a patient can lay in bed with a nurse taking measurement of the patient's blood pressure, administering oxygen, or writing notes to be saved to the patient's inventory.

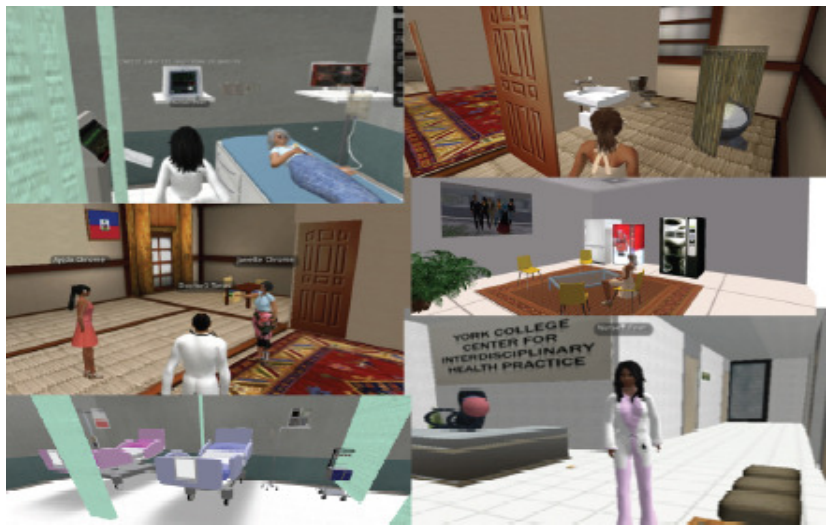

Figure 2: A virtual hospital and a patient's house with doctors and patients interacting with each other.

We also built a virtual two-story resident house with typical accommodations such as a bathroom, a bedroom, and a living room. It serves as a patient residence with furniture and living conditions that fit to the descriptions designated by our faculty for case studies.

## 4. FINDINGS

The approach of building a virtual learning environment in professional education promoted interdisciplinary collaboration and was technically feasible and financially cost-effective.

- It enhanced interdisciplinary collaboration among faculty for promoting quality care and patient safety.

  Our faculty were actively involved in the design and development of the project. They communicated through email and face-to-face meetings. The vignettes they designed were based upon typical cases they use in their classrooms. The opportunity of designing cases that can be adopted into different disciplines invited interesting discussions among faculty in a broad research scope and allowed them to share visions across subject domains. For instance, some faculty stressed the importance of patient-centered care for healthcare improvement; while other faculty noted it was important to design teaching strategies to promote student-centered instruction. Although from two different disciplines (i.e., nursing education and teacher education), they agreed the patient/student-centered approach could serve patients/students more efficiently - it allows practitioners to design healthcare/teaching strategies that include factors such as race, age, culture, learning styles, etc.

- We designed the project with a shared goal: enhance higher order thinking skills among students that can be transferred to real clinical settings (e.g., decision-making, problem solving, and analytical/critical thinking);

  Observing how domain experts make decisions is a process of making thinking visible. It is consistent with the cognitive apprenticeship model, which is difficult to widely apply in traditional classrooms due to physical/economic limitations. As Collins [23] states, too little attention is paid to the reasoning and strategies that experts employ when they acquire knowledge or put it to work to solve complex or real-life tasks.

The International Journal of Multimedia & Its Applications (IJMA) Vol.3, No.1, February 2011

In our 3D teaching vignettes, avatars demonstrate what domain experts think and how they deal with real-life like situations. The higher-order thinking processes externalized serve as expert mental models for students to observe. For instance, in two of our storylines, an alcoholic wife and a Haitian elderly patient have different history of illness. They also have different cultural and economic backgrounds. As providing adaptive patient care is crucial for improving patient care and cutting costs in the long run, all these factors have to be considered to make a sound medical decision. Also getting the students immersed in the real-life like setting encourage them to identify with the patients and motivate them to actively seek solutions as an expert with mindfulness. To achieve this, we zoomed in to the patients showing vital signs of illness (sweating, shaking, etc) and made the experts think out aloud and provide treatment based upon specific situations (Figure 3).

We plan to assess learning outcomes and student attitudes towards such a virtual learning environment in a study in the near future. Bloom's [24] taxonomy of learning will be used as our evaluation model regarding student attitude, knowledge, and skills. More specifically, we will assess students' attitudes, motivation, and self-perception of ability through instructor observations and student self-assessment. Domain knowledge and skills will be evaluated through role-playing, quizzes, debriefing notes, or self-reflection.

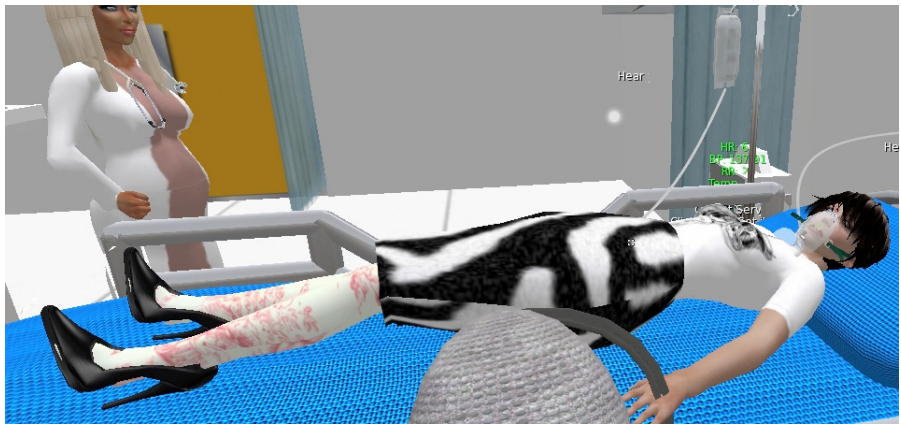

Figure 3. Nurse treating an alcoholic female in the hospital

- Developing a virtual learning environment is technically feasible and financially cost effective.

    We found building these teaching vignettes helped us understand the advantages and challenges of promoting user-centered instruction through 3D simulation. It is technically and financially feasible to create a 3D virtual learning environment with a quick turnaround time. Many virtual items can be purchased online at a low price and customized based upon specific needs. Once the "stage" is set up, the "actors" can start to perform and develop dynamic stories on the same platform with little extra cost. A video capture tool is set up so that the stories, that the avatar controllers carried out, can be recorded as raw movie clips for editing later. Sometimes, the same story needs to be played several times from different angels or views so that the scenes can be stitched together as a more coherent movie.

- Such a viable virtual learning environment can be easily customized to situate students in a new learning context.
106

We believe our framework can be easily applied to other subject domains, such as classroom management in Teacher Education, helping instructors design teachable moments in a 3D simulated world. As the land and virtual rooms has been purchased and set up, all one needs to get started is to obtain some virtual items in a school setting such as desks, chairs, blackboard, teachers and students, etc (Figure 4). Such items cost little or nothing, as they are common items available in the virtual market place. Therefore, researchers and educators can focus on the design of the lessons with little initial cost and technical difficulties.

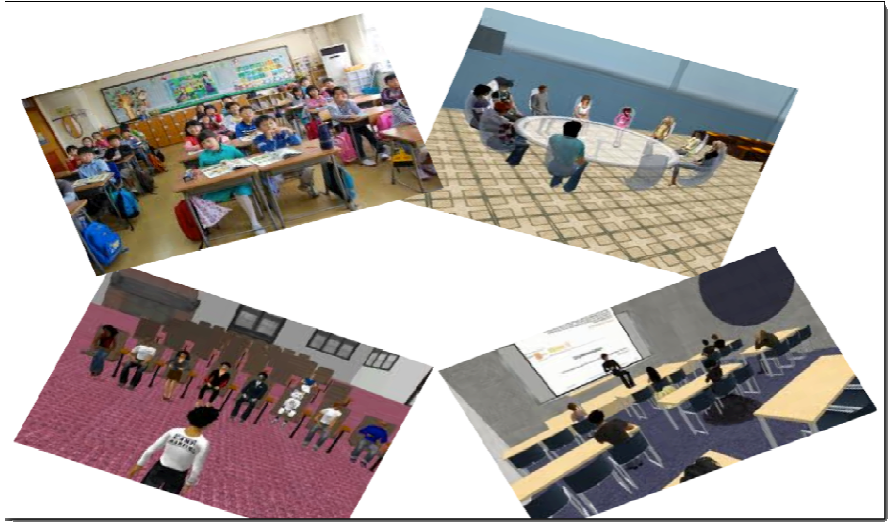

Figure 4. A mockup screenshot of a virtual classrooms to prepare our students for student teaching in our virtual learning environment

- Creating 3D virtual learning environment still faces many challenges

There is still a learning curve to be able to navigate in a virtual world. Users need orientation and practices to control their avatars. They also need to learn how to interact with other avatars and the virtual environment. For instance, they need to grab a virtual scripting item to make an avatar to escort another avatar so that they can walk hand-in-hand. Also they sometimes need to concatenate a few gestures together to make a new behavior, such as seizure or fainting. Those actions require no cautious effort in real life. But in Second Life, users have to create their own customized usually behaviors.

Also, the costs of running these games with high-speed Internet connection and powerful computers may discourage some students and faculty from participation. We found some textures in the virtual environment, such as wallpaper, trees, or avatars, take longer to load. Sometimes we cannot add complex virtual items, as there is a limit to the number of virtual items we can put to a virtual land. These constraints offset some of the affordances the virtual reality technology can offer.

## **5. CONCLUSION**

We developed 3D teaching vignettes in a social networking environment based upon the collaboration among seven faculty members that are from five disciplines. The 3D teaching vignettes modelled domain experts demonstrating how they provide adaptive interventions based upon individual patient's needs. These video clips, combining with other activities such as online reflections/discussions, could allow our students to reach out to domain experts, peers, and observe how experts solve real-life like problems. We found such an approach encouraged interdisciplinary collaboration. It was a viable approach to design an interactive learning

The International Journal of Multimedia & Its Applications (IJMA) Vol.3, No.1, February 2011

environment with lower costs. We expect to use the platform to encourage students to engage in role-play with peers in the virtual hospital in the near future. We hope this learning-by-doing approach can be applied to other settings, such as classroom management in Teacher Education. This will encourage faculty across the campus to participate in designing effective virtual learning environment as an alternative teaching tool in the classroom.

## ACKNOWLEDGMENTS

This project was undertaken as a collaborative project at York College - The City University of New York. The project team consisted of the following faculty: Dana Fusco from Academic Affairs/Teacher Education, Beverly Horowitz from Occupational Therapy, Susan Glodstein from Nursing, Joanne Lavin from Nursing, John Graffeo from Physician Assistant, Robert Duncan from Psychology, Beth Rosenthal from Social Work, and Xin Bai from Educational Technology/Teacher Education. Support for this project has been provided by City University of New York's Workforce Development Initiative and Incentive grant programs. We want to thank all involved in this project.

**Authors**

**Xin Bai** is an assistant professor of Educational Technology in Department of Teacher Education at York College, City University of New York. She earned her doctorate in Instructional Technology & Media from Teachers College at Columbia University. Her research focuses on educational games, simulations, intelligent tutoring systems, e-learning, and ubiquitous learning. Her work is built on the research done on cognitive science, artificial intelligence, and educational technologies.

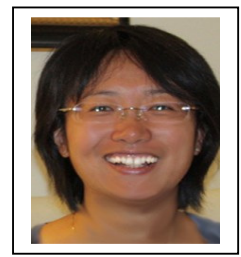

**Dr. Dana** Fusco serves as Acting Associate Dean for the School of Health & Behavioral Sciences at York College and Associate Professor in Teacher Education. As Dean, she has worked to support interdisciplinary curricular, research and other projects across the School's health, social and behavioral sciences. She oversees the college's CUE programs, outcomes assessment, Title III, testing, and the York Early College Academy. For the past twenty years, Dr. Fusco's research has focused on youth development and afterschool programs. Her work in this area has led to increased national and international recognition. Recently she was the Keynote Speaker at the History of Youth Work conference in Minnesota and presented at the International Conference on Youth Work and Youth Studies in Glasgow. She was invited to join the national panel of leaders in youth work, the Next Generation Coalition. She has authored dozens of peer-reviewed articles, wrote and produced a video documentary, and most recently is working on an edited volume to be published by Routledge next year entitled, Advancing Youth Work: Current Trends, Critical Questions.

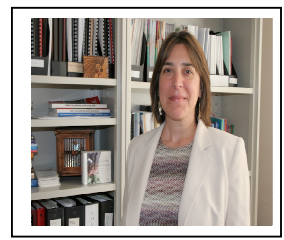